\pdfoutput=1

\documentclass[12pt,a4paper,notitlepage,oneside]{book}

\usepackage{alltt}
\usepackage[T1]{fontenc}
\usepackage[utf8]{inputenc}
\usepackage[english]{babel}

\usepackage{amsmath, amsthm, amssymb, mathtools}
\usepackage{dsfont}
\usepackage[cal=cm]{mathalfa}
\numberwithin{equation}{section}
\usepackage{chngcntr}
\counterwithout{equation}{section}
\usepackage{subfigure}
\usepackage{setspace} 
\usepackage{graphicx}
\usepackage{adjustbox}
\usepackage[usenames,dvipsnames]{xcolor}

\usepackage{float} 

\usepackage{booktabs}

\usepackage{todonotes} 
\usepackage{url}
\usepackage{pdflscape} 
\usepackage{framed} 
\usepackage{enumitem} 
\usepackage[iso,swedish]{isodate}

\usepackage[round,colon,authoryear]{natbib}

\usepackage{abstract_def}

\usepackage{titling}
\usepackage{authblk}
\usepackage{bm}
\title{Iterated filtering methods for Markov process epidemic models}
\author[1]{Theresa Stocks}
\affil[1]{Department of Mathematics, Stockholm University, Sweden}
\date{\vspace{-5ex}}

\begin{document}
\maketitle

\framebox{
  \begin{minipage}{0.9\linewidth}
    This manuscript is a preprint of a Chapter to appear in the
    \textit{Handbook of Infectious Disease Data Analysis}, Held, L.,
    Hens, N., O'Neill, P.D. and Wallinga, J. (Eds.). Chapman \& Hall/CRC,
    2018. Please use the book for possible citations.
  \end{minipage}
}\\[2ex]

\begin{chapabstract} 
  \small{
\noindent 
Dynamic epidemic models have proven valuable for public health decision makers as they provide useful insights into the understanding and prevention of infectious diseases. However, inference for these types of models can be difficult because the disease spread is typically only partially observed e.g. in form of reported incidences in given time periods. This chapter discusses how to perform likelihood-based inference for partially observed Markov epidemic models when it is relatively easy to generate samples from the Markov transmission model while the likelihood function is intractable. The first part of the chapter reviews the theoretical background of inference for partially observed Markov processes (POMP) via iterated filtering. In the second part of the chapter the performance of the method and associated practical difficulties are illustrated on two examples. In the first example a simulated outbreak data set consisting of the number of newly reported cases aggregated by week is fitted to a POMP where the underlying disease transmission model is assumed to be a simple Markovian SIR model. The second example illustrates possible model extensions such as seasonal forcing and over-dispersion in both, the transmission and observation model, which can be used, e.g., when analysing routinely collected rotavirus surveillance data. Both examples are implemented using the R-package \texttt{pomp}  \citep{kingNgu} and the code is made available online. 
}
  
\end{chapabstract}
\thispagestyle{empty}
\thispagestyle{empty}
\clearpage
\setcounter{page}{1}

\setcounter{chapter}{0}

\chapter[Iterated filtering methods for POMP]{Iterated filtering methods for Markov process epidemic models}

\section{Introduction}
In this chapter we describe how iterated filtering methods \citep{ionides06, Ionides2015}
can be used to analyze available infectious disease outbreak data in the form of time series.
The centerpiece of these methods is the assumption that the outbreak data can be modeled as a  noisy and only partially observed realization of a disease transmission process that is assumed to be a Markov process \citep{kingNgu}.
The general inference approach is to (i) formulate a suitable Markovian transmission process; (ii) connect the data to the transmission process via some suitable observation process; (iii) use iterated filtering to perform inference for the model parameters.  
The inference  method presented here is likelihood-based. It is designed for models where it is relatively easy to draw samples from the Markov process compared to evaluating its transition probabilities. The iterated filtering algorithm is, among others, implemented in the \texttt{R} package \texttt{pomp} \citep{kingNgu} which spans a wide collection of simulation, inference and model selection methods for partially observed Markov processes (POMP). Other simulation-based   inference methods for this model class are  simulated moments \citep{sm}, synthetic likelihood \citep{sl}, non-linear forecasting  \citep{nlf} or  Bayesian approaches such as  approximate Bayesian computations \citep{abc, liu}  and  particle MCMC \citep{pmcmc}. However, at present  iterated filtering methods are the only currently available, frequentist, full-information, simulation-based methods for POMP models \citep{Ionides2015}.\\
In this chapter we focus  on the ``simplest'' Markovian SIR trasmission model and  describe in some detail how to fit this model to outbreak data consisting of the number of newly reported cases aggregated over  time intervals, e.g.\ weeks. However, the methods can be easily extended to  more complicated settings, several of which will be discussed here. \\ The chapter is structured as follows:  Section \ref{likelihood} gives a short overview about likelihood-based inference and describes some of its general challenges.  In Section \ref{inference} we introduce the  model class of partially observed Markov processes and  explain how to formulate and evaluate the likelihood of such models with particle filters. In Section \ref{iterated_filtering} we present the iterated filtering algorithm  and  demonstrate its use by a simple example  in Section \ref{examplesir}. In Section \ref{complications} we discuss possible extensions of this example and, in Section \ref{rota_example}, illustrate  how the method can be applied to a real-world problem which accommodates most of the  complications mentioned. Both examples are accompanied by source code and instructions which can be found online at \cite{theresa_git}.  We finish the chapter by outlining  advantages and disadvantages of the method presented and point the interested reader to further literature, see Section \ref{comments}. 
\section{Likelihood-based inference}
\label{likelihood}
We start by outlining the key aspects of likelihood-based inference that will be relevant to us. This is an extensive topic and for readers new to this area we refer to e.g.\  \cite{Pawitan2001} or \cite{held}  for a comprehensive overview on likelihood-based inference. 

\subsection{The likelihood approach}
The idea behind likelihood-based inference is the following. Suppose we have data in the form of  a sequence of $N$ observations $\bm{y}^*_{1:N}$ at times $t_1,\dots ,t_N$ and a model for the data $f(\bm{y}_{1:N};\bm{\theta})$ where $f$ is typically a probability mass or density function parametrized by  a vector of parameters denoted by $\bm{\theta}$. In our context $\bm{y}^*_{1:N}$ might, for example, represent the weekly number of newly reported cases over a certain period of time  and $\bm{\theta}$ might contain the  parameters of a suitable Markovian epidemic  transmission model. In order to calibrate the model to our observations we would like to find the elements of the  parameter vector $\bm{\theta}$ for which our observations are most likely under the chosen probability model. In other words, we would like to maximize this function  $f$ with respect to $\bm{\theta}$ evaluated at the data  $\bm{y}^*_{1:N}$. This translates to optimizing the function
\begin{align*}
\mathcal{L}(\bm{\theta};\bm{y}^*_{1:N})&=f(\bm{y}^*_{1:N};\bm{\theta}).
\end{align*} 
The function $\mathcal{L}$ is called the \textit{likelihood function} and in the following we will  suppress its dependence on the data and simply write $\mathcal{L}(\bm{\theta})$ for convenience. The parameter vector which maximizes this function  is called the \textit{maximum likelihood estimate} (MLE) and is given as 
\begin{align}
\hat{\bm{\theta}} =\underset{\bm{\theta}\in \bm{\Theta}}{\arg\max} \mathcal{L}(\bm{\theta}),
\label{mle}
\end{align}
where $\bm{\Theta}$ is the parameter space containing all possible sets of parameters. \\
For many applications it is often more convenient to work with the \textit{log-likelihood function} 
\begin{align*}
l(\bm{\theta})=\log \mathcal{L}(\bm{\theta}).
\end{align*} 
This transformation often simplifies optimization, but does not change the location of the MLE since the natural logarithm is a monotonically increasing function.
\subsection{Practical challenges}
In principle, the optimization problem in (\ref{mle}) looks rather straightforward, however, there are a number of challenges in practice. Firstly, the evaluation of $\mathcal{L}$  can be difficult because the function might not be available in closed form. Secondly, even if evaluation is possible, it might be very hard to derive  the first and higher order derivatives of $\mathcal{L}$ analytically or even numerically which are needed for numerical optimization methods, cf.\ \cite{Nocedal1999}.  In this case we need derivative-free optimizers. Those optimizers might impose other problems, for example when the likelihood can only be approximated stochastically, e.g.\ by Monte Carlo methods \citep{Robert2004}. In that case standard deterministic derivative-free optimizers fail. All the  problems  mentioned above occur for the model class at hand because the likelihood is a complex integral. In the following sections we will give more rigorous details about the problem and introduce a method which gets around these challenges. The problems mentioned arise in the specific setting of our model formulation. In addition,  there are some other general challenges with likelihood-based inference in a statistical context. 
Usually, the point estimate we obtain from maximum likelihood estimation is not very meaningful by itself unless we also quantify the uncertainty of the estimate. 
 One way to solve this problem in a likelihood setting  is to  construct confidence intervals for the parameters by e.g.\ calculating the profile log-likelihood for each parameter of interest and invert Wilks' likelihood ratio test to get the desired intervals \citep{Wilks1938}. Another very common challenge is that often there exist multiple local maxima and the optimization algorithm can get stuck in one of these and not return the global maximum. It is therefore important to use a wide range of starting values for the numerical algorithm  in order to improve chances that the global maximum is reached.  Moreover, it might very well happen that the   maximum is not unique because the surface of the likelihood function has ridges. In that case confidence intervals are a good way to quantify the parameter range.  
In the following, we introduce and formulate the likelihood  of partially observed Markov processes and explain how to tackle all of the issues mentioned above.
\section{Inference for partially observed Markov processes (POMP)}
\label{inference}
In the literature partially observed Markov processes  are also known as hidden state space models or stochastic dynamical systems. The main assumption is that at discrete points in time we observe some noisy aspect of the true  underlying Markov process which is itself often continuous in time. In the following we  formulate the likelihood of a POMP, give an idea why standard methods to find the MLE do not apply here and finally describe how iterated filtering methods overcome these problems. In the following, the exposition and notation  is adopted from \cite{kingNgu} and the materials and tutorials in  \cite{King2017}. 
\begin{figure}[h!]
\begin{center}
\includegraphics[width=12cm]{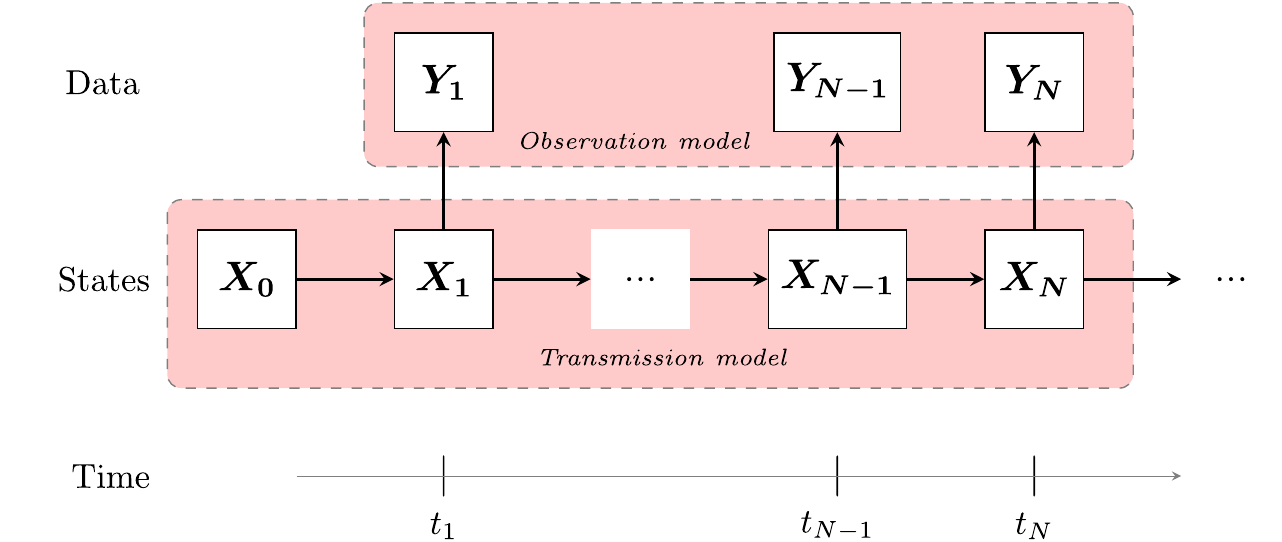}
\end{center}
\caption{A partially observed Markov process where $\bm{Y}_n, n=1,\dots,N$, denotes the observations at time $t_n$, which depend on the state of the transmission process $\bm{X}_n:= \bm{X}(t_n;\bm{\theta})$ at that time. }
\label{pomp}
\end{figure}\\

\subsection{Likelihood of a POMP}
A POMP consists of two model components, (i) an unobserved Markov process $\{\bm{X}(t;\bm{\theta}): t \geq 0 \}$, which can be discrete or continuous in time,  and (ii) an observation model which describes how the data collected at discrete points in time $t_1,\dots ,t_N$, is connected to the transmission model. For notational convenience, we write $\bm{X}_n=\bm{X}(t_n;\bm{\theta})$
and $\bm{X}_{0:N}=(\bm{X}_0,\dots,\bm{X}_N)$. In our application, the process $\bm{X}_{0:N}$ describes the dynamics of the disease spread, e.g.\ in the case of the simple Markovian SIR model $\bm{X}_n= (S(t_n),I(t_n), R(t_n))'$  counts the number of susceptible, infectious and removed individuals at time $t_n$.
Let $\bm{Y}_n$ denote the random variable counting the observations at time $t_n$ which depend on the state of the transmission process $\bm{X}_n$ at that time, cf.\  Figure \ref{pomp}. 
Our data, $\bm{y}^*_{1:N} = (\bm{y}^*_{1} , \dots,\bm{y}^*_{N})$, are then
modeled as a realization of this observation process.  Depending  on how many aspects of the disease dynamics we observe, $\bm{y}^*_{1:N}$ can be either a univariate or multivariate time series. Assuming that the observable random variable  $\bm{Y}_n$ is independent of all other variables given  the state of the transmission process $\bm{X}_{n}$, the joint density of the states and the observations is defined as the product of the   one-step transmission density, $f_{\bm{X}_n|\bm{X}_{n-1}}(\bm{x}_n|\bm{x}_{n-1};\bm{\theta})$, the  observation density, $f_{\bm{Y}_n|\bm{X}_{n}}(\bm{y}_n|\bm{x}_{n};\bm{\theta})$, and the initial density $f_{\bm {X}_0}(\bm{x}_0;\bm{\theta})$ as 
\begin{align*}
f_{\bm{X}_{0:N},\bm{Y}_{1:N}}(\bm{x}_{0:N},\bm{y}_{1:N};\bm{\theta})=&f_{\bm{X}_0}(\bm{x}_0;\bm{\theta})\\
&\times \prod_{n=1}^{N}f_{\bm{X}_n|\bm{X}_{n-1}}(\bm{x}_n|\bm{x}_{n-1};\bm{\theta})f_{\bm{Y}_n|\bm{X}_{n}}(\bm{y}_n|\bm{x}_{n};\bm{\theta}).
\end{align*} 
Hence, the likelihood of the parameter vector can be written as the marginal density for a sequence of observations $\bm{Y}_{1:N}$ evaluated at the data $\bm{y}^*_{1:N}$ as

\begin{align}
\mathcal{L}(\bm{\theta})&=f_{\bm{Y}_{1:N}}(\bm{y}^*_{1:N};\bm{\theta})=\int f_{\bm{X}_{0:N},\bm{Y}_{1:N}}(\bm{x}_{0:N},\bm{y}^*_{1:N};\bm{\theta})d\bm{x}_{0:N}.
\label{lik}
\end{align}
The dimension of the integral in (\ref{lik}) depends on the number of compartments of the Markov process and the number of observations, so this integral is usually high-dimensional and, except for the simplest situations, cannot be solved analytically. The method we will present in the following sections uses the property that it is fairly easy to draw samples from the density $f_{\bm{X}_n|\bm{X}_{n-1}}$ rather than evaluating the function. Because the distribution  $f_{\bm{X}_0}$ of the initial state of the Markov process is usually not known,  one practical way to deal with this is to  fix the initial values of the system  at some  reasonable values or,  treat them  as parameters we would like to estimate, see Section \ref{complications} for further discussion.  
 \subsection{Evaluation of the likelihood}
 \label{evaluation}
 Before addressing the question of  how to maximize the likelihood in (\ref{lik}) in an efficient way, we have to first think about how we actually evaluate $\mathcal{L}$ for a given parameter vector $\bm{\theta}$. 
In the special case that the underlying transmission model is deterministic and the initial values are given,  $\bm{X}_n=\bm{x}_n(\bm{\theta})$ is a non-random function of $\bm{\theta}$ for each $n$, hence $f_{\bm{X}_n|\bm{X}_{n-1}}$and $f_{\bm{X}_0}$   are point masses. 
The likelihood  equation (\ref{lik}) then reduces to 
\begin{equation}
\ell(\bm{\theta})=\log \mathcal{L}(\bm{\theta})= \sum_{n=1}^N \log f_{\bm{Y}_n|\bm{X}_{n}}(\bm{y}^*_n;\bm{x}_{n}(\bm{\theta}),\bm{\theta}).
\end{equation} 
So given that we know the distribution of our observation model, evaluation of the log-likelihood corresponds to  summing over the logarithm of the observation density given the states of the deterministic ``Markov process'' at each point in time. Maximum likelihood estimation then  reduces to a classical numerical optimization problem for a non-linear  function of the parameter vector and for optimization we can use a derivative-free optimizer e.g.\ the  Nelder-Mead method \citep{nelder}. 
However, in general  evaluating (and optimizing) the function $\mathcal{L}$ is unfortunately not as straightforward because the transmission model is not deterministic.
Rather, it is a non-trivial problem because of the high dimension of the integral.  In the following we describe how to evaluate Equation (\ref{lik})  in a general way with particle filters.
 \subsection{Particle filter}
One standard approach to solve a high dimensional integral  as in Equation (\ref{lik}) is  to approximate the integral by Monte Carlo methods. However, as it will turn out, for the model class at hand, this approach is highly inefficient in  practice. Nevertheless, we will demonstrate this approach first, in order to introduce the reader to the problem.  Readers unfamiliar with  Monte Carlo methods might want to consult the relevant literature, e.g.\ \cite{Robert2004, pfilter}. \\
The basic idea is that the likelihood in Equation (\ref{lik}) which can also be written as the integral 
\begin{equation}
\mathcal{L}(\bm{\theta})=\int 
 \prod_{n=1}^{N}f_{\bm{Y}_n|\bm{X}_{n}}(\bm{y}_n|\bm{x}_{n};\bm{\theta})f_{\bm{X}_{0:N}}(\bm{x}_{0:N};\bm{\theta})d\bm{x}_{0:N}
\end{equation}
can be approximated using the Monte Carlo principle as
\begin{align*}
\mathcal{L}(\bm{\theta})&\approx \frac{1}{J}\sum_{j=1}^{J}\prod_{n=1}^{N}f_{\bm{Y}_n|\bm{X}_{n}}(\bm{y}^*_{n}|\bm{x}_{n,j};\bm{\theta}),
\end{align*}
where $\{\bm{x}_{0:N,j},j=1,\dots ,J\}$ is a sample drawn from $f_{\bm{X}_{0:N}}(\bm{x}_{0:N};\bm{\theta})$.
This means that if we  generate trajectories by simulating from the Markov process, we just have to evaluate the density of the data given the realizations and then average in order to  obtain a Monte Carlo approximation of the likelihood.\\
It turns out, however, that with this approach the  variability in the approximation of the  likelihood is very high, because the way of  proposing trajectories is entirely unconditional of the data. This means that in practice this method is very inefficient, because a lot of simulations are necessary in order to obtain an estimate of the likelihood that is precise enough to be useful for parameter estimation or model selection. The longer the time series, the worse the problem gets \citep{King2017}. As it turns out, a better idea is to parametrize the likelihood in (\ref{lik})  as 
 \begin{align}
\mathcal{L}(\bm{\theta})&=\prod_{n=1}^N \mathcal{L}_{n|n-1}(\bm{\theta}),
\end{align}
where $\mathcal{L}_{n|n-1}(\bm{\theta})$ is the conditional likelihood, given as 
\begin{align}
\mathcal{L}_{n|n-1}(\bm{\theta})&= f_{\bm{Y}_n|\bm{Y}_{1:n-1}}(\bm{y}^*_{n}|\bm{y}^*_{1:n-1};\bm{\theta}) \nonumber\\&= \int f_{\bm{Y}_n|\bm{X}_{n}}(\bm{y}^*_{n}|\bm{x}_n;\bm{\theta})f_{\bm{X}_n|\bm{Y}_{1:n-1}}(\bm{x}_{n}|\bm{y}^*_{1:n-1};\bm{\theta})d\bm{x}_n, \label{conlik} 
\end{align}
with the convention that $f_{\bm{X}_1|\bm{Y}_{1:0}}=f_{\bm{X}_1}$.\\
In the following we explain how this integral can be efficiently approximated with resampling techniques which are the basis of particle filter methods.
Since $\bm{X}_n$  is a  Markov chain it follows from the Chapman Kolmogorov equation that
\begin{align}
f_{\bm{X}_n|\bm{Y}_{1:n-1}}&(\bm{x}_{n}|\bm{y}^*_{1:n-1};\bm{\theta})= \label{prediction} \\ 
& \int f_{\bm{X}_n|\bm{X}_{n-1}}(\bm{x}_{n}|\bm{x}_{n-1};\bm{\theta})f_{\bm{X}_{n-1}|\bm{Y}_{1:n-1}}(\bm{x}_{n-1}|\bm{y}^*_{1:n-1};\bm{\theta})d\bm{x}_{n-1}\nonumber.
\end{align}
From Bayes' theorem it also follows that
\begin{align}
f_{\bm{X}_n|\bm{Y}_{1:n}}(\bm{x}_{n}|\bm{y}^*_{1:n};\bm{\theta})&=f_{\bm{X}_n|\bm{Y}_{n},\bm{Y}_{1:n-1}}(\bm{x}_{n}|\bm{y}^*_{n},\bm{y}^*_{1:n-1};\bm{\theta})\nonumber \\&=\frac{f_{\bm{Y}_n|\bm{X}_{n}}(\bm{y}^*_{n}|\bm{x}_{n};\bm{\theta})f_{\bm{X}_n|\bm{Y}_{1:n-1}}(\bm{x}_{n}|\bm{y}^*_{1:n-1};\bm{\theta})}{\int f_{\bm{Y}_n|\bm{X}_{n}}(\bm{y}^*_{n}|\bm{x}_{n};\bm{\theta})f_{\bm{X}_n|\bm{Y}_{1:n-1}}(\bm{x}_{n}|\bm{y}^*_{1:n-1};\bm{\theta}) d\bm{x}_n},\label{filtering}
\end{align}
where we use the fact that $\bm{Y}_n$ only depends on $\bm{X}_n$.
The distribution $f_{\bm{X}_{n}|\bm{Y}_{1:n-1}}$ is called the \textit{prediction distribution} and 
$f_{\bm{X}_{n}|\bm{Y}_{1:n}}$  is called the \textit{filtering distribution} at time $t_n$. The idea now works as follows. Assume we have a set of $J$ points
$\{\bm{x}^F_{n-1,j}\}_{j=1}^J $,  in the following referred to as \textit{particles},   from the filtering distribution at time $t_{n-1}$. Equation (\ref{prediction}) then implies that we obtain a sample $\{\bm{x}^P_{n,j}\}$ from the prediction distribution at time $t_n$  if we just simulate from  the Markov model    
\begin{equation}
\bm{x}^P_{n,j} \sim f_{\bm{X}_n|\bm{X}_{n-1}}(\ \cdot \ |\bm{x}^F_{n-1,j};\bm{\theta})\quad \text{with}\quad j=1,\dots , J.
\label{simulating}
\end{equation}
Equation (\ref{filtering}) in turn  tells us that resampling from $\{\bm{x}^P_{n,j}\}$ with weights proportional to 
\begin{equation}
\omega_{n,j}=f_{\bm{Y}_n|\bm{X}_{n}}(\bm{y}^*_{n}|\bm{x}^P_{n,j};\bm{\theta})
\label{weight}
\end{equation}
 gives us a sample from the filtering distribution at time $t_n$.  By the Monte Carlo principle it follows from (\ref{conlik}) that 
\begin{align*}
\mathcal{L}_{n|n-1}(\bm{\theta})& \approx \frac{1}{J}\sum_{j=1}^J f_{\bm{Y}_n|\bm{X}_{n}}(\bm{y}^*_{n}|\bm{x}^P_{n,j};\bm{\theta}),
\end{align*}
where $\bm{x}^P_{n,j}$ is approximately drawn from $f_{\bm{X}_n|\bm{Y}_{1:n-1}}(\bm{x}_{n}|\bm{y}^*_{1:n-1};\bm{\theta})$.
In order to obtain the full likelihood we have to iterate  through the data, alternating  between simulating (\ref{simulating})  and resampling (\ref{weight}) in every time step, until we reach  $n=N$ so that
\begin{equation}
\ell(\bm{\theta})= \log \mathcal{L}(\bm{\theta})= \sum_{n=1}^N \log \mathcal{L}_{n|n-1}(\bm{\theta}).
\label{eval_pfilter}
\end{equation}
This method to evaluate the likelihood is called a \textit{sequential Monte Carlo} algorithm or \textit{particle filter}   \citep{ Kitagawa1987,pfilter, Arulampalam2002}. The method  is implemented as the \texttt{pfilter} function   in the \texttt{pomp} package \citep{kingNgu} and  returns a stochastic estimate of the likelihood which can be shown to be unbiased \citep{DelMoral1996}. \\
It can happen that at some time point a very unlikely particle is suggested. In practice, if the conditional likelihood of this particle  is below  a certain tolerance value, then that particle is considered to be uninformative.
If,  at some time point, the conditional likelihood of every particle is below this chosen tolerance a \textit{filtering failure} occurs. When a failure occurs,  re-sampling is omitted, the conditional likelihood at that time point is set  equal to the tolerance and the iteration through the data set continues, hence the  time point at which the filtering failure occurred is taken to contain no information. In general, filtering failures are an implication that the model and data might not be consistent \citep{King2017}.\\
The variability of the approximation in Equation (\ref{eval_pfilter}) can be reduced by increasing the number $J$ of particles, however, the variability will usually not vanish completely. 
This might create  problems for standard optimizers which assume that the likelihood is evaluated deterministically. A better choice in this case is to use  stochastic optimizers  such as the iterated filtering method that we present in the section below.
\section{Iterated filtering methods}
\label{iterated_filtering}
Iterated filtering  is a simulation-based method  to find the MLE which takes advantage of the structure of POMP models and particle filters. It was first introduced  by \cite{ionides06} and further improved in \cite{Ionides2015}. The iterated filtering method uses the property that it is easy to simulate from $f_{\bm{X}_n|\bm{X}_{n-1}}$ while the likelihood is not tractable directly.
The basic idea is that a particle filter is applied to a model in which  the parameter vector for each particle is following a random walk.
As iterations progress, the intensity of the perturbations is successively reduced ("cooling").
It can be shown that the algorithm converges towards the MLE \citep{Ionides2015}. At present  this method is the only simulation-based frequentist approach that uses the full information contained in the data. Moreover, iterated filtering methods have been able to solve likelihood-based inference problems for infectious disease related questions which were computationally intractable for available Bayesian methods \citep{Ionides2015}.
\subsection{Algorithm}
\label{mif_algorithm}
In the following we present the pseudocode for iterated filtering as implemented in the \texttt{mif2} function in the \texttt{R} package \texttt{pomp} and explain how to draw samples from $f_{\bm{X}_n|\bm{X}_{n-1}}$. 
\\
\newline
\textbf{Pseudocode iterated filtering (\texttt{mif2}) }cf.\ \cite{Ionides2015}\\
{\footnotesize{\textbf{Input:}  
 Simulators for $f_{\bm{X}_0}(\bm{x}_0;\boldsymbol{\theta})$ and $f_{\bm{X}_n|\bm{X}_{n-1}}(\bm{x}_n| \bm{x}_{n-1}; \boldsymbol{\theta})$;  
 evaluator for $f_{\bm{Y}_n|\bm{X}_n}(\bm{y}_n|\bm{x}_n;\boldsymbol{\theta})$;  
data $\bm{y}^*_{1:N}$ \\
\textbf{Algorithmic parameters:}
\# of iterations $M$;  
\#  of particles $J$;  
initial parameter swarm $\{\boldsymbol{\theta}^0_j, j=1,\dots,J\}$;  
perturbation density $h_n(\boldsymbol{\theta}|\varphi;\sigma)$;  
perturbation scale $\sigma_{1{:}M}$ \\
\vspace{0.1cm}
\textbf{Procedure:}\\
1. $\quad$ For $m$ in $1{:} M$\\
 2. $\quad\quad\quad$ $\boldsymbol{\theta}^{F,m}_{0,j}\sim h_0(\ \cdot \ |\bm{\theta}^{m-1}_{j}; \sigma_m)$ for $j$ in $1{:} J$\\
 3. $\quad\quad\quad$ $\bm{X}_{0,j}^{F,m}\sim f_{\bm{X}_0}(\ \cdot \ ; \bm{\theta}^{F,m}_{0,j})$ for $j$ in $1{:} J$\\
 4. $\quad\quad\quad$ For $n$ in $1{:} N$\\
 5. $\quad\quad\quad\quad\quad$ $\bm{\theta}^{P,m}_{n,j}\sim h_n(\ \cdot \ |\bm{\theta}^{F,m}_{n-1,j},\sigma_m)$ for $j$ in $1{:} J$\\
 6. $\quad\quad\quad\quad\quad$ $\bm{X}_{n,j}^{P,m}\sim f_{\bm{X}_n|\bm{X}_{n-1}}(\ \cdot \  |\bm{X}^{F,m}_{n-1,j}; \bm{\theta}^{P,m}_{n,j})$ for $j$ in $1{:} J$\\
 7. $\quad\quad\quad\quad\quad$ $w_{n,j}^m = f_{\bm{Y}_n|\bm{X}_n}(\bm{y}^*_n| \bm{X}_{n,j}^{P,m} ; \bm{\theta}^{P,m}_{n,j})$ for $j$ in $1{:} J$\\
 8. $\quad\quad\quad\quad\quad$ Draw $k_{1{:}J}$ with $P(k_j=i)=  w_{n,i}^m\Big/\sum_{u=1}^J w_{n,u}^m$\\
 9.  $\quad\quad\quad\quad\quad$ $\bm{\theta}^{F,m}_{n,j}=\bm{\theta}^{P,m}_{n,k_j}$ and $\bm{X}^{F,m}_{n,j}=\bm{X}^{P,m}_{n,k_j}$ for $j$ in $1{:} J$\\
 10. $\quad\quad\quad$ End For\\
 11. $\quad\quad\quad$ Set $\bm{\theta}^{m}_{j}=\bm{\theta}^{F,m}_{N,j}$ for $j$ in $1{:} J$\\
 12. $\quad$ End For\\
 \vspace{0.5cm}
\textbf{Output:}  
Final parameter swarm, $\{\bm{\theta}^M_j, j=1,\dots,J\}$ }}\\
In the algorithm the initial parameter values are perturbed (line 2) by a perturbation density where its  standard  deviation $\sigma_m$ is a decreasing function of $m$. The Markov process is initialized (line 3) as a draw from the initial density dependent on the proposed parameter vector. What follows in lines 4 - 10 is a particle filter as described in the section above with the only difference being that the parameter vector is stochastically perturbed in every iteration through the data. The 
$M$ loop repeats the  particle filter with decreasing perturbations and throughout the algorithm the 
superscripts $F$ and $P$ denote filtering and prediction distribution, respectively. The algorithm returns the best guess of the parameter swarm after $M$ iterations. In the \texttt{R} package \texttt{pomp} the point estimate that the function \texttt{mif2} returns is the mean of the parameter swarm.
\newline
To generate realizations from $f_{\bm{X}_n|\bm{X}_{n-1}}$  we can use the Gillespie algorithm \citep{gila}. Given the current state of the system, the algorithm simulates the waiting time of the next event and updates the number of individuals in each compartment and the overall  time is incremented  accordingly.  The whole procedure is repeated until a pre-defined stopping time is reached. In the case of constant per capita transition rates, the simulation of every individual event gives us a complete and detailed  history of the process, however, it is usually a very time-consuming task for systems with large population and  state space, because of the enormous number of events that can take place. 
  As a way to speed up such simulations an approximate simulation method can be used, the so called \textit{$\tau$-leap algorithm} which is based on the Gillespie algorithm \citep{gil}.  
The $\tau$-leap algorithm holds all rates constant in a small time interval $\tau$ and simulates the numbers of events that will occur in this interval, then updates all state variables, computes the transition rates again and the procedure is repeated until the stopping time is reached \citep{taul,into}. 
Given the total number of jumps, the number of individuals leaving any of the states by any available route during a  time interval $\tau$ is then  multinomially distributed. Note that the simulation time step is  not identical with the time scale the observation process evolves on.
Both simulation algorithms are conveniently implemented in the \texttt{pomp} package as the functions \texttt{gillespie.sim} and \texttt{euler.sim}.
\section{Iterated filtering for SIR model given incidence data}
\label{examplesir}
\subsection{The problem of interest}
Typical  infectious disease data  collected by  public health authorities often consists of reported incidences in given time periods. Other important characteristics of an epidemic such as recovery times of the individual  or contact network are not observed. The method presented is a useful tool to analyze routinely collected surveillance data because it gives insights into the mechanism of  disease spread which is crucial if one wants  to e.g.\ assess the risk of emerging pathogens or evaluate the impact of control measures such as vaccination. \\
In this section we describe how to carry out  inference for the parameters of a POMP where the underlying disease transmission model is assumed to be a simple Markovian SIR model, given  that we observe the number of newly reported cases aggregated by week. 
We use the algorithm presented in Section \ref{iterated_filtering} for a simulated data set and illustrate how the algorithm performs. \\
In order to keep things simple, in the specific example of this section, we assume that the time of reporting coincides with the time of infection.  Of course, this assumption does not hold for all diseases but the
method presented here can be easily adopted to other settings, e.g.\ when the time of  reporting coincides with the time of removal.\\
The implementation of the following example is made available at \cite{theresa_git}.

\subsection{Formulation of a POMP model}
We will first formulate a Markov transmission model and in a second step relate the data to the transmission model via some observation model which then gives us $f_{\bm{Y}_n|\bm{X}_{n}}$. \\
\newline
\textbf{Transmission model} 
As transmission model we choose  a stochastic SIR model among a closed population of $\mathcal{N}$ individuals where  
$\bm{X}(t)=(S(t), I(t), R(t)) $ denotes the number of susceptible, infectious and recovered individuals at time $t$. Individuals are mixing homogeneously and $\beta$ is the average number of infectious contacts an infectious individual has per time unit. Furthermore, we assume that the time an individual is infectious is exponentially distributed with mean $\gamma ^{-1}$. 
We will now fromulate this in a way which makes understanding of the code  in \cite{theresa_git} easier. So  let  $\{N_{AB}(t): t \geq 0\}$  denote a stochastic process which counts the number of individuals which have  transitioned from compartment $A$ to compartment $B$ during the time interval $[0,t)$ with $A, B \in \mathcal{X}$, where $\mathcal{X}=\{S, I, R\}$ contains all model compartments. The infinitesimal increment probabilities of a jump between compartments fully specify the continuous-time Markov process describing the transmission dynamics. With the notation above,  $\Delta N_{AB}(t)=N_{AB}(t+\tau)-N_{AB}(t)$ counts the number of individuals changing compartment in an infinitesimal time interval $\tau>0$. Thus, 
 \begin{align}
\mathbb{P}[\Delta N_{SI}(t)=1 |\bm{X}(t)]&=  \beta I(t) S(t)\mathcal{N}^{-1}\tau +o(\tau) \quad \nonumber \\
\mathbb{P}[\Delta N_{IR}(t)=1 |\bm{X}(t)]&= \gamma I(t)\tau  +o(\tau). \quad\label{jump}
\end{align}
Moreover, the state variables and the transition probabilities (\ref{jump}) are  related in the following way:
\begin{align}
\Delta S(t)&= -\Delta N_{ SI}(t)  \nonumber\\
\Delta I(t)&= \Delta N_{ SI}(t)-\Delta N_{ IR}(t) \label{states}\\
\Delta R(t)&= \Delta N_{IR}(t). \nonumber
\end{align}
\newline
\textbf{Observation model}
We now relate the transmission model to our observations of the weekly number of newly reported cases.  With the notation above
 the true number of newly infected individuals accumulated  in each observation time period $(t_{n-1},t_{n}], n \in \{1,2,\dots,N\}$  is given as 
\begin{equation}
H(t_n)=N_{SI}(t_{n})-N_{SI}(t_{n-1}).\label{obs_stoc}
\end{equation}
To incorporate the  count nature of the observations a natural assumption is  to model the actual reported cases as  realizations of a Poisson-distributed random variable with a given time-dependent mean. 
\noindent The number of recorded  cases $Y_n$ within a given reporting interval $(t_{n-1},t_{n}],$ is then 
\begin{align}
Y_n &\sim \text{Pois}\left(  H(t_n)\right).
\label{poisson}
\end{align}
This distribution then corresponds to $f_{\bm{Y_n|\bm{X}_n}}$ which is easy to evaluate. One interpretation of this choice of observation noise is to account for uncertainty in classification of cases, including false positives. 

\subsection{Inference}
\begin{figure}[H]
\begin{center}
\includegraphics[width=10cm]{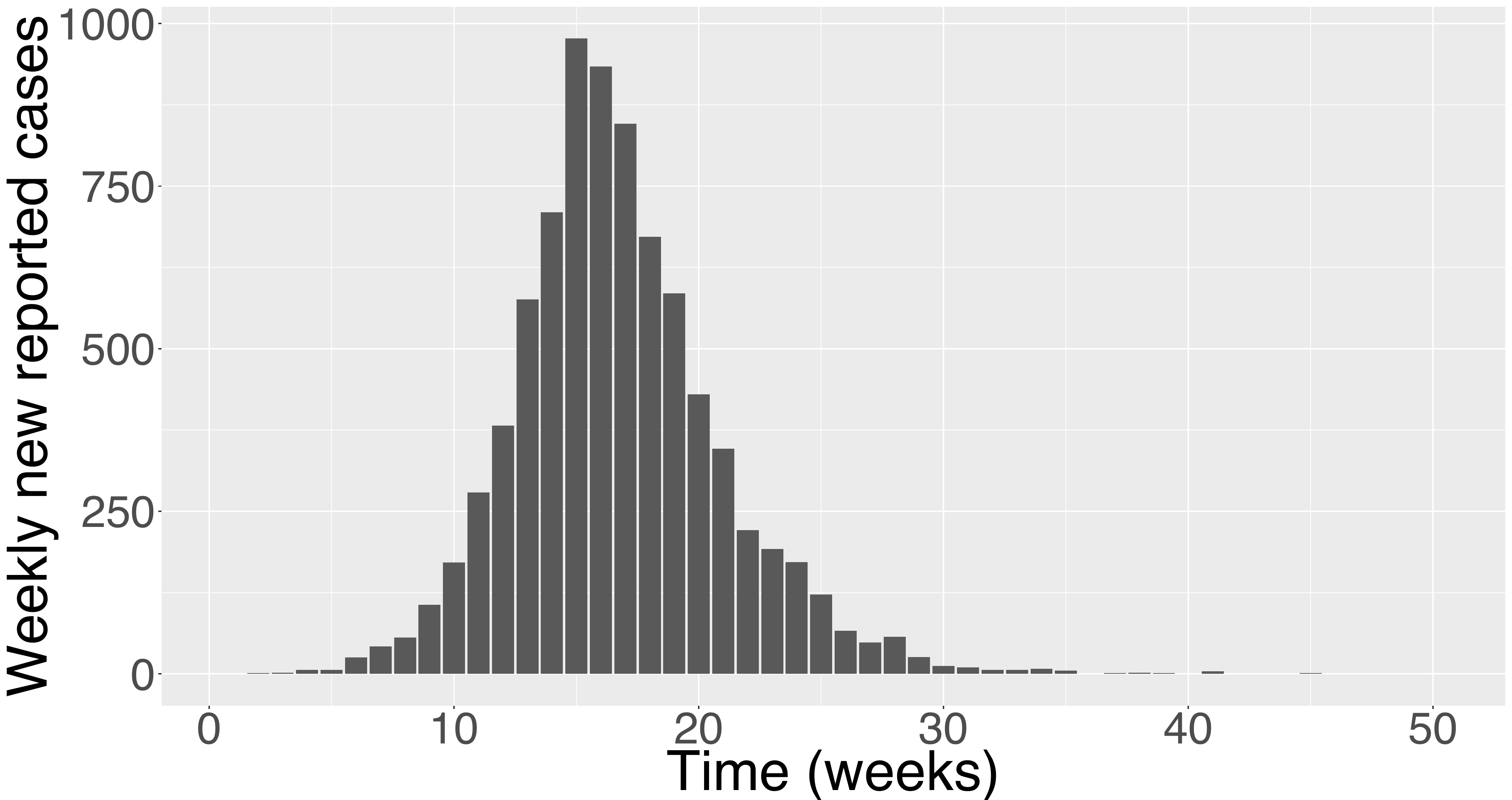}
\end{center}
\caption{Weekly number of newly reported cases  over time, for a simulated set from the POMP  model with an underlying  Markovian SIR transmission model as defined in Equations  (\ref{states}) and (\ref{poisson}).}
\label{sim_data}
\end{figure}

\begin{figure}[H]
\begin{center}
\includegraphics[width=12cm]{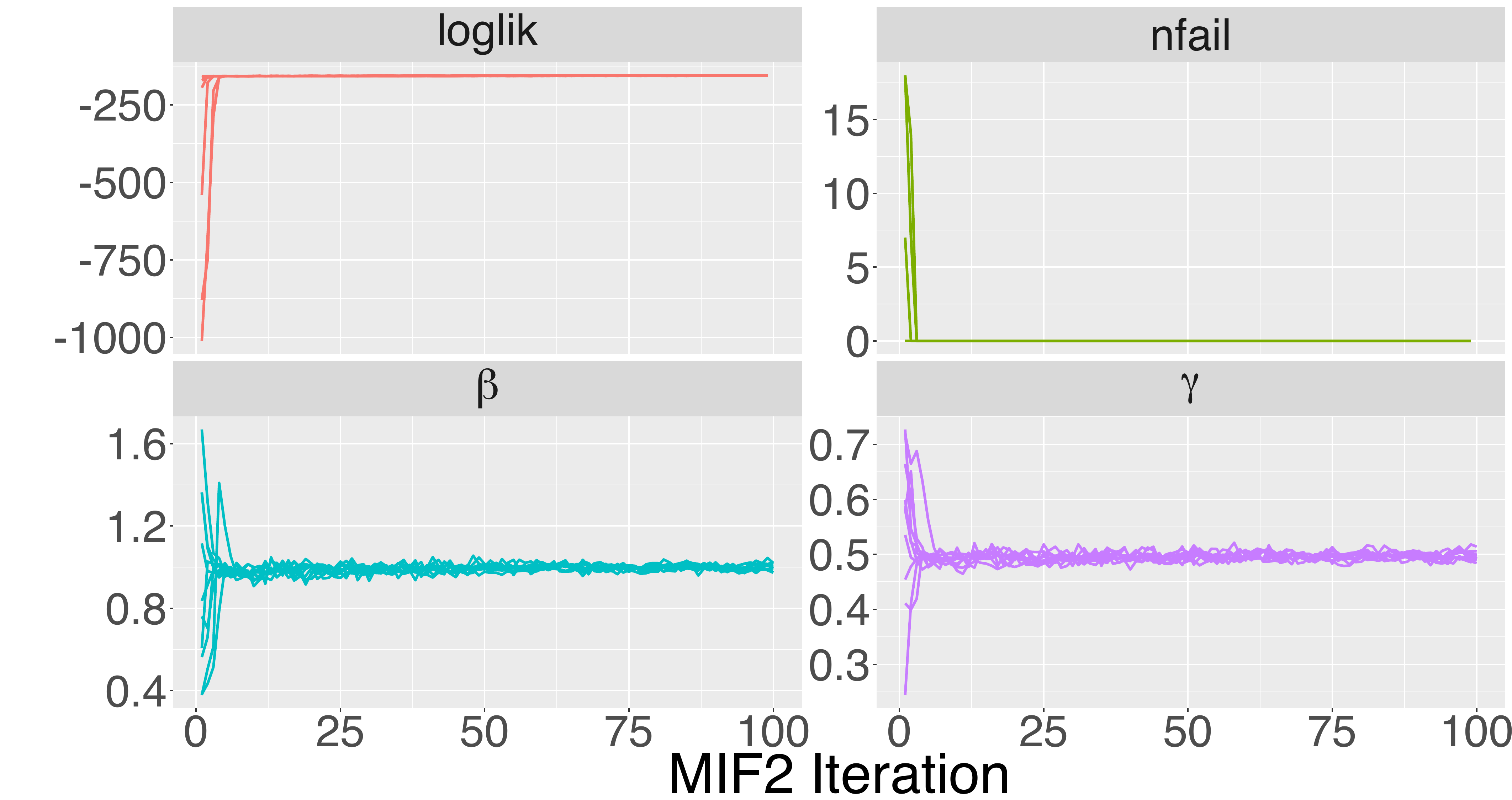}
\end{center}
\caption{Diagnostic plot of the iterated filtering algorithm for 100 iterations for simulated  incidence data. Shown is the evolution of the log-likelihood (loglik), the number of filtering failures in each \texttt{mif2} iteration (nfail) and parameter estimates for parameters $\beta$ and $\gamma$ per iteration for 10 trajectories with random starting values drawn from a hypercube.}
\label{mif}
\end{figure}

\begin{figure}[H]
\centering     
\subfigure[Profile log-likelihood for $\beta$]{\label{profile_b}\includegraphics[width=55mm]{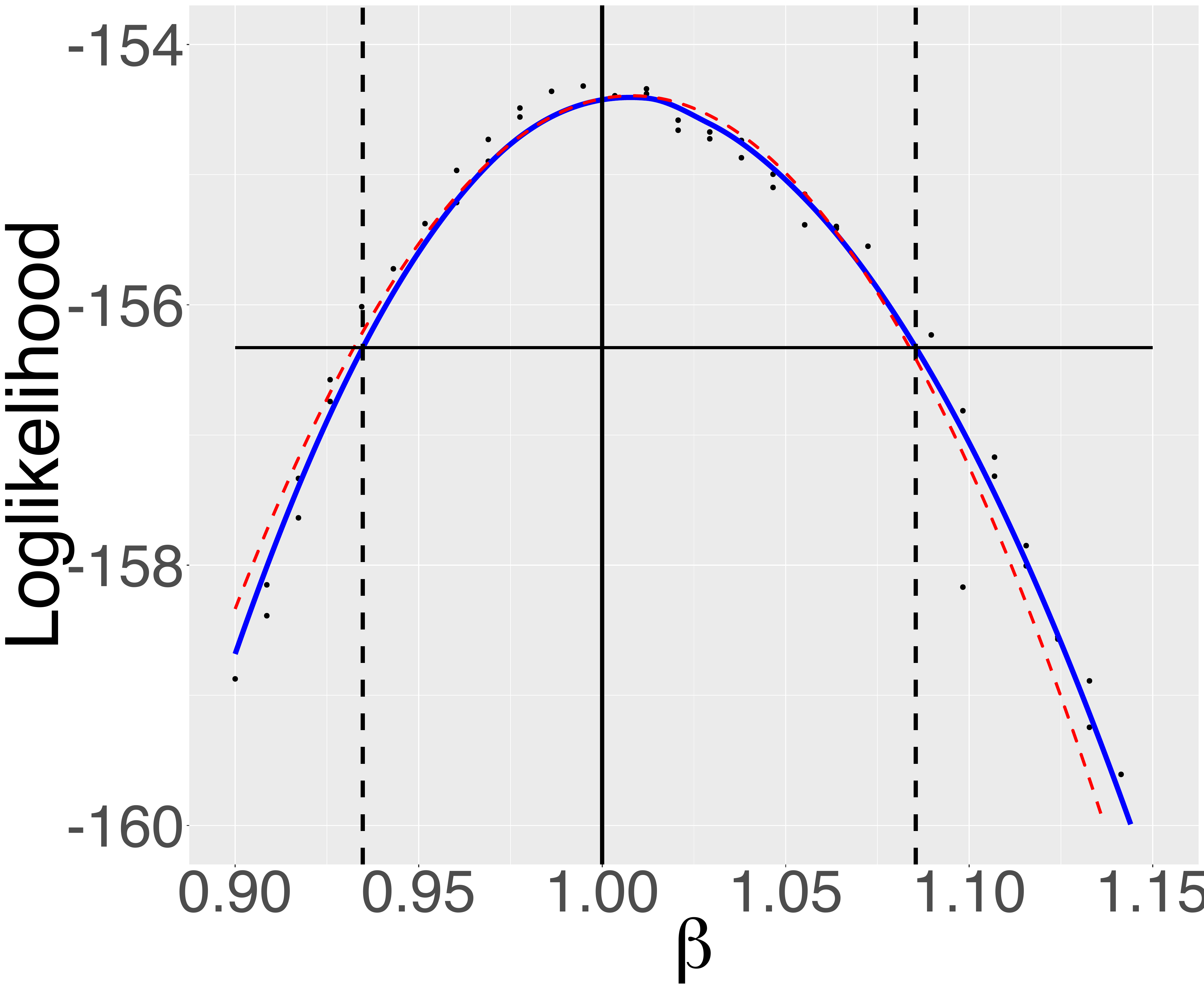}}
\subfigure[Profile log-likelihood for $\gamma$]{\label{profile_g}\includegraphics[width=55mm]{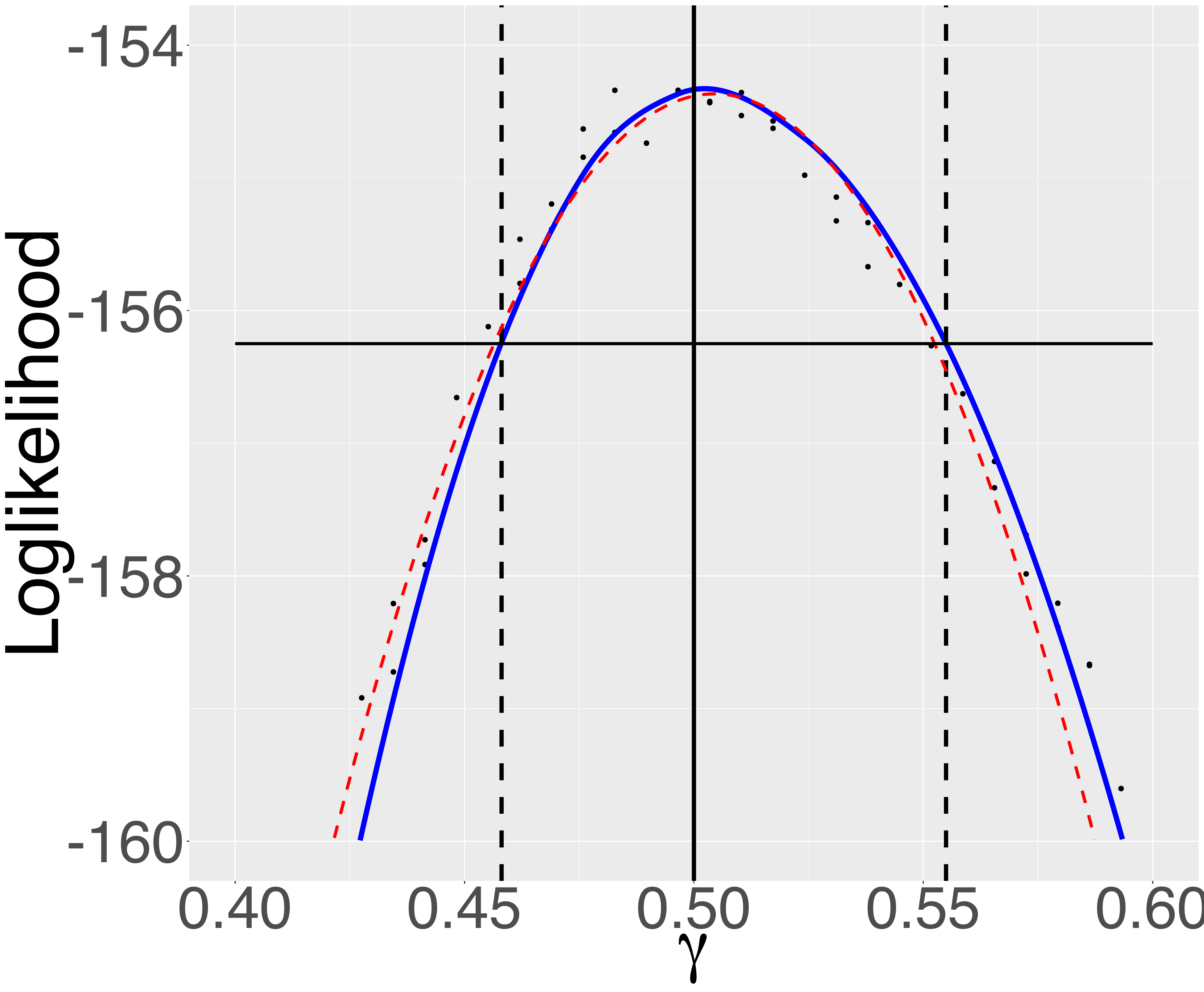}}
\caption{The smoothed profile likelihoods (solid curves) for both parameters of interest  where the solid vertical lines show the true values and the dashed vertical lines contain the corresponding MCAP 95\%
confidence interval. The quadratic approximation in a neighborhood
of the maximum is shown as the dashed curves.
}\label{profile}
\end{figure}

In the following we perform inference for the two epidemiological parameters $\beta $ and $\gamma$ on a set of simulated data from the model where we apply the iterated filtering algorithm presented in Section \ref{iterated_filtering}. 
\\
\newline
\textbf{Data} We consider a closed population with $\mathcal{N}=10\ 000$ individuals of whom one is initially infectious and the rest are susceptible. A realization from the model presented above  with $\beta=1$ and $\gamma=0.5$ for $N=50$ weeks where $t_n=n$ for $n=1,\dots,N$ was simulated. Figure \ref{sim_data} shows the number of weekly reported cases where it is assumed that all cases are reported.
\noindent \textbf{Implementation details} We use the  iterated filtering algorithm as implemented in the function \texttt{mif2} in the R package \texttt{pomp} to infer  the epidemic parameters $\beta$ and $\gamma$ assuming the initial values and total population size were known. We run the algorithm for  $M=100$ iterations and $J=500$ particles. As starting values for the fitting algorithm we use 10 parameter constellations drawn uniformly from a hypercube which contains biologically meaningful parameter values for $\beta$ and $\gamma$. With this procedure we address the potential problem of local maxima in the optimization.
 In general, if  all searches converge to the same MLE  for starting values drawn at random from a hypercube this increases the chances that a global maximum has been found \citep{King2017}.  
For  the perturbation of the parameters we choose a random walk with an initial standard derivation of $0.02$ for both parameters. As cooling scheme we use geometric cooling which means that  on the n-th \texttt{mif2} iteration, the relative perturbation intensity is $k^{n/50}$ where we chose $k=0.05$. That is, after 50 iterations the perturbation is only half of the intensity compared to the first iteration.  
In each iteration  the variance of the random walk is successively reduced so the log-likelihood of the perturbed model gradually approaches the log-likelihood of  the model we are interested in. However, after a finite number of iteration steps, the log-likelihoods of the two models, the one with and the one without perturbed parameters,  are not identical. To get around this issue a particle filter evaluation of the \texttt{mif2} model output using Equation (\ref{eval_pfilter}) is necessary. Hence, for each  of the 10 \texttt{mif2} outputs we obtain we run 10 particle filters,  each with 1000 particles. From the multiple particle filter evaluation we can then calculate the average  log-likelihood and the standard error of the Monte Carlo approximation  for every parameter set. Consequently, we  choose the parameter constellation of the 10 possible with the highest average log-likelihood as the maximum likelihood  estimate.  
In order to quantify the associated uncertainties of the parameter estimates based on our observations we calculate the 95 \% confidence intervals for each parameter.  For this  we construct the profile likelihood of each parameter and apply the Monte Carlo adjusted profile (MCAP) algorithm \citep{Ionides2017}. This recently developed methodology accounts for the presence of Monte Carlo error in the likelihood evaluation and adjusts the width of the confidence interval  accordingly. 
This procedure might seem overly complicated at first sight because, intuitively, the parameter swarm we obtain as an output from the iterated filtering algorithm should contain some measure of uncertainty. However, due to particle depletion, which is the situation when only very few different particles have
significant weight \citep{Doucet2011}, the information about  the local shape of the likelihood surface contained in the particles is not very reliable. The profile likelihood is much more robust in this case since it relies on multiple independent \texttt{mif2} searches \citep{King2017}.\\
\newline
\textbf{Results} The inference results are visualized in Figures \ref{mif} and \ref{profile}. As shown in Figure \ref{mif} the randomly drawn starting values for both parameters are converging towards the true values at the same time as the likelihood increases and the number of filtering failures disappears. Both 95\% confidence intervals constructed from the profile log-likelihoods cover the true parameters as shown in Figure  \ref{profile}.

\subsection{Practical advice} 
\label{tips}
In the \texttt{R} package \texttt{pomp} a partially observed Markov process together with observations in the form of a uni- or multivariate time series can be represented via a so called \texttt{pomp} object. The whole package's functionality such as simulation and inference is centered around this object. Depending on which functions one would like to use, the model can be implemented by specifying some or all model components. In order to help identify possible error sources that can occur while applying the algorithm we now describe a few practical hints.\\
The first step after constructing the \texttt{pomp} object is to simulate realizations from the model in order to check that the code is working and the model gives the expected output.  A good idea at this stage is  to also construct the corresponding deterministic skeleton of the stochastic Markov transmission model in order to have a simplified version of the model and to also generate trajectories and compare if the mean of the stochastic model corresponds to the deterministic version. Another benefit of simulating from the model in general is that it  helps  to identify a range of likely parameters which generate realizations that resemble the data.
Before the data comes into play and the actual fitting starts we find it very helpful to first test the algorithm with simulated data. For this a realization of the model is generated, then the known set of parameters which was used to generate the realization is evaluated with the particle filter. This way we make sure that the core function in the iterated filtering method is working. 
Then the algorithm can be applied to the simulated set and it should give back the values used for simulation. This procedure also  helps understanding  how the parameter estimators  are correlated and if they are identifiable.\\
 Now, let us move to the data.
As a last step before applying the  iterated filtering algorithm to our observations  we recommend to fit the data first to the simpler, deterministic  model in order to get a feeling for how the parameters behave and identify possible problems with the likelihood surface etc. 
Then, when the \texttt{mif2} is finally used it makes sense to choose likely values as starting values for the parameters we want to estimate first. However,  in a last step a global search should be carried out which means that random starting values are chosen  and, ideally, then all searches converge to the same MLE regardless of their starting values. In practice, the likelihood is often ridged so some searches can get stuck in local maxima. That is why it is important to use many initial values from many different starting values. Another common problem in practice is that parameter values are suggested which are very unlikely which in turn leads to filtering failures. In that case it can be useful to add stochasticity to the model which accounts for model misspecifications or unmodelled disease characteristics. \\
Working with the \texttt{mif2} algorithm can be a time consuming task because the method is simulation-based. This is the price we pay for not needing to be able to evaluate the density function of the Markov model directly. That is  why we also want to give some hints of how to speed up the procedure to some extent. The first idea is to parallelize the code so that multiple searches can be carried out at the same time. This, in turn,  poses some challenges in preventing repetition of expensive computations and trying to ensure reproducibility and -- we present one way of doing this  in the manual \cite{theresa_git}, more can be found in the tutorials at \cite{King2017}.
In order to ensure that the information about the likelihood surface is 
 continually improved it is useful to keep a record of all parameter vectors estimated with the \texttt{mif2} function, together with the computed likelihood for each parameter set,
e.g.\ by saving the obtained results in a CSV file \citep{King2017}.
 Moreover, it is sometimes worth accepting approximation errors by choosing the simulation time step in the $\tau$ leaping algorithm to be not too small, which can lead to significant gains in computational speed; for a short discussion on  numerical solutions based on discretizations in the context of POMP models cf.\ \cite{breto}. 
\section{Extensions}\label{complications}
In the previous section we used a  simple model to explain the concept of iterated filtering  in order to facilitate understanding. However, reality is often way more complicated than this toy example. In the following we discuss some important complications of infectious disease data and how they can be included in POMP models.\\
\newline
\textbf{Underreporting} Underreporting is a common problem appearing in surveillance data and is usually the more pronounced the less severe the disease is. It arises for different reasons e.g.\ it can result from asymptomatic cases, cases where individuals do not consult the doctor, cases that are not identified as such (misdiagnosis) or  cases which just get lost in the reporting system \citep{Gibbons2014}. One way to include underreporting in a POMP model is to  change the observation model to 
\begin{align*}
Y_n &\sim \text{Pois}\left(  \kappa \cdot H(t_n)\right).
\end{align*}
where  $\kappa \in (0,1]$ is the reporting rate which then also can be estimated.\\
\newline
\noindent \textbf{Seasonality}
For many diseases such as childhood diseases or vector borne diseases the transmission rate $\beta$ is not constant but varies through time. This might be due to social aggregations of the host such as in  daycare institutions and schools which are closed during summer, changes in environment that influence the biology of vector populations or changes in weather such as temperature or precipitation. One way to introduce this complexity into our models is through seasonal  forcing  of the transmission rate so that the number of infectious contacts changes through time as  $\beta(t)$. Following \cite{kee}, one possible choice is
\begin{equation}
\beta(t)= \beta \left(1+\rho\cos\left(\frac{2\pi}{w} t +\phi\right)\right), 
\label{forcing}
\end{equation}
where $\rho\in [0,1] $ is the amplitude of the forcing,  $2\pi/w \in \mathbb{R}^+$ is the period of the forcing  and $\phi \in [0,2\pi]$  is the phase shift parameter. 
With this choice of forcing function the parameter  $\beta $ can be interpreted as the average transmission rate of an individual which varies between $(1-\rho)\beta$ and $(1+\rho)\beta$ during the forcing period. \\
A more flexible way to choose the forcing function is through splines. If $s_i(t)$, $i = 1,\dots, k$ is a periodic B-spline basis,  then
we can for example define
\begin{align*}
\beta(t)= \sum_{i=1}^k b_i s_i(t).
\end{align*}
Variation of the coefficients $b_i$ then gives a wide variety of shapes for $\beta(t)$. For the reader new to the theory of splines we recommend \cite{Stoer2002}. A comprehensive overview of the problems of seasonal forcing can be found in \cite{kee}.\\
\newline
\textbf{Overdispersion} A common phenomenon in calibration of models from data is the  presence of greater variability in the  set of observations  than would be expected  under the assumed model, cf. e.g.\ \cite{McCullagh1989, breto}. Reasons for this include  heterogeneities in the population, model misspecifications or unmodelled processes which we either cannot quantify or are very hard to measure in any way. 
This phenomenon is called overdispersion and we can account for it in two different ways.  On the one hand,  we can introduce overdispersion into the observation model
by changing the observation distribution from the Poisson distribution  to a distribution where  the variance can be adjusted separately from the mean. In the case of count data, a natural distribution would be the negative binomial distribution, so 
\begin{align}
Y_n &\sim \text{NegBin}\left( H(t_n), \frac{1}{\psi}\right),
\label{nbin}
\end{align}
with  $H(t_n)$ being the true number of   accumulated incidences per time unit  $(t_{n-1},t_{n}]$  and $\text{NegBin}(\mu,1/\psi) $ with $ \psi>0$  denotes the negative binomial distribution with mean $\mu$ and variance $\mu+\psi\mu^2$. 
Another  choice could be the truncated normal distribution where the results are  rounded  to integers. \\
However, in outbreak data, we often see fluctuations which  are clearly not only due to variation in data collection,  but due to phenomena not captured by the model. 
In the toy example in Section \ref{examplesir}, we have accounted for stochasticity in the underlying system by assuming that individuals move between classes at random times. However, for large population sizes  the stochastic system approaches the deterministic system  and, hence, the role of randomness decreases as the population size increases; for details cf.\ e.g.\ \cite{dar}. 
One way to account for overdispersion in the transmission model is to  stochastically perturb the transmission rate  by
multiplying a continuous-time white noise    process $\xi(t)$ which fluctuates around the value one with the transmission rate.  
In \cite{breto} it is shown that by choosing the corresponding integrated noise process $\Gamma(t)$ in a way  such that its increments are independent, stationary, non-negative  and unbiased the Markov property is retained. 
One suitable example for such a process is  a L\'evy process  with
\begin{align*}
\xi(t)&=\frac{d}{dt}\Gamma(t),\quad \text{where marginally } \quad  \Gamma(t+\tau )-\Gamma(t)\sim \text{Gamma}\left(\frac{\tau}{\sigma^2},\sigma^2\right),
\end{align*}
and where $\tau /\sigma^2$ denotes the shape and $\sigma^2$  the age-independent scale parameter with corresponding mean $\tau$ and variance $\tau\sigma^2$.  
The parameter $\sigma^2$ is called the \textit{infinitesimal variance} parameter \citep{karlin}. We can include this into the model by letting  the  transmission rate  be
\begin{equation}
\beta(t)=\beta\cdot \xi(t). \label{gammawnoise}
\end{equation} 
More details about this kind of overdispersion can be found in \cite{breto,Breto2011}.\\
\newline
\textbf{Structured populations} To make mathematical models for disease spread more realistic different kinds of host heterogeneities can be introduced \citep{kee}. For example age structure, spatial heterogeneity or varying risk structures of the individuals might play an important role in disease transmission and  can be  easily accommodated in a POMP.  \\
\newline
\textbf{Covariates}  It can be interesting  to investigate what impact a vector-valued covariate process $\{Z(t): t \geq 0\}$ has on the disease dynamics. 
 That might  e.g.\ be environmental covariates such as temperature,  precipitation, or demographic information such as numbers of births and deaths over time. Inclusion of such a covariate vector is unproblematic in a POMP setting because the densities $f_{\bm{X}_0}$, 
$f_{\bm{X}_n|\bm{X}_{n-1}}$ and $f_{\bm{Y}_n|\bm{X}_n}$ can depend on the observed process $\{Z(t):t \geq 0\}$ \citep{kingNgu}.  If one is e.g.\ interested in investigating the influence of a specific covariate on the transmission rate one  could check the model's capability to fit the data with and without this covariate and compare the results. 
\\
\newline
\textbf{Initial values}
In our toy example of Section \ref{examplesir}  we assumed that the distribution of the initial values of the Markov process, $f_{\bm{X}_0}$, is known. Besides some special cases that could arise e.g.\ in some experimental situations, this distribution is, however, not known or observable. If the disease we observe is  endemic in the population and does not vary significant over time, it might make sense to assume that the initial values originate from the stationary distribution of the system. If this is not the case we can treat the value of $\bm{X}_0$ as a parameter to be estimated. In this case, $f_{\bm{X}_0}$ is a point mass, the
concentration of which depends on $\bm{\theta}$ \citep{kingNgu}.\\
\newline
\textbf{Missing data} Missing values in outbreak data are a common complication.  When using the \texttt{R} package \texttt{pomp}, the \texttt{pomp} object  can handle missing values  if the observation model probability density function is  implemented so as to deal with them appropriately. 
One way is to  construct the  observation density  so it  returns   e.g.\ a log-likelihood  of $0$ for the missing data entries, for implementational details c.f.\ the FAQ in \cite{King2017}.

\section{Rotavirus example}

\label{rota_example}
\begin{figure}[h!]
\begin{center}
\includegraphics[width=11.5cm]{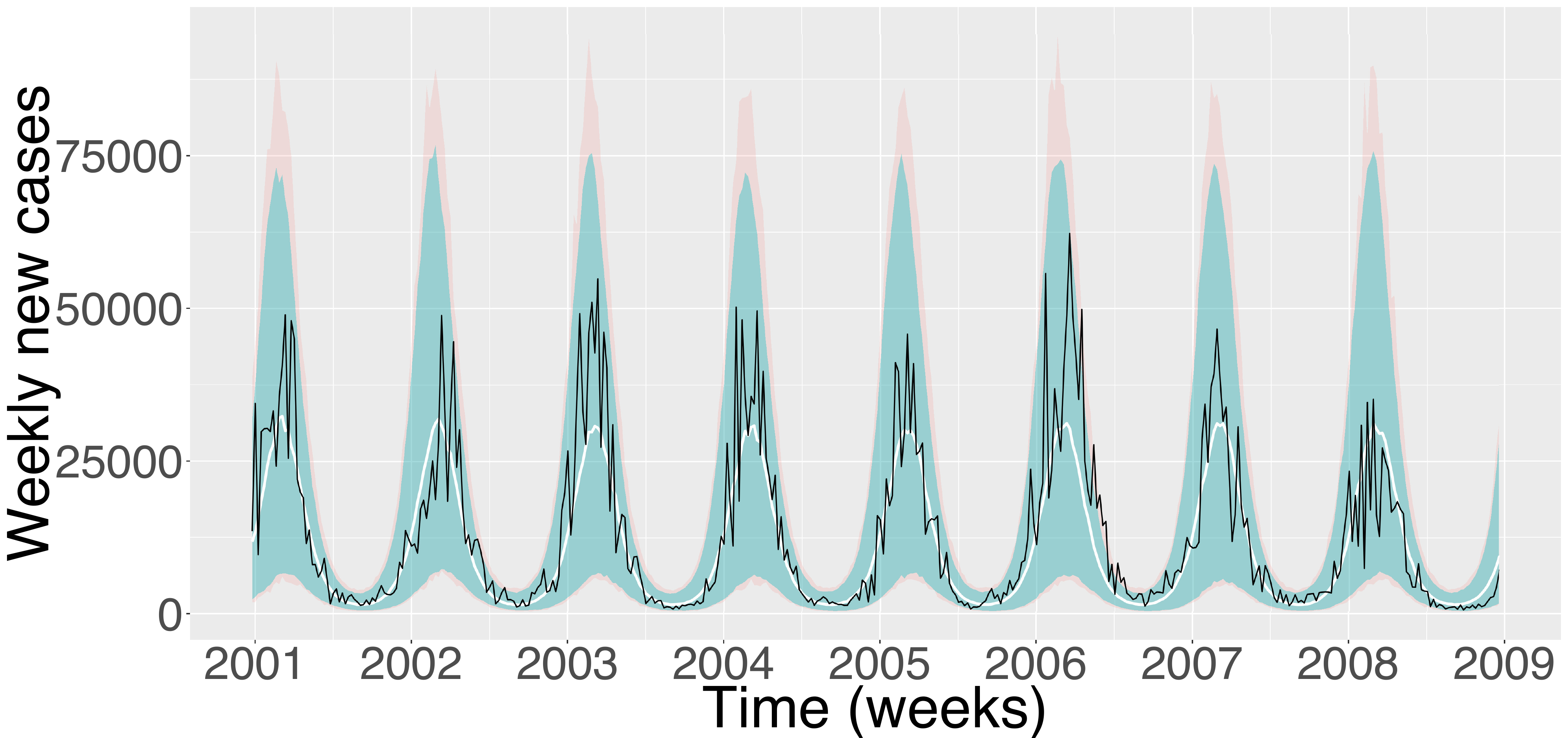}
\end{center}
\caption{The 95\% prediction interval (light shading) for  1000 realizations of the  model for rotavirus evaluated at the maximum likelihood estimate for the simulated case report data (solid back line) and the median (solid white line). Furthermore, the 95 \% prediction interval of these 1000 realizations for only the transmission model  is shown (darker shading). }
\label{rota}
\end{figure}

After presenting a  simple toy example  as a proof of concept, we now  give a short illustration of a real-world problem which accommodates some of the  complications presented in the previous section and was analyzed with iterated filtering. A detailed description of the model and inference results can be found in \cite{Stocks2017}. \\
The data which were analyzed in that paper are the weekly reported number of new laboratory-confirmed rotavirus cases among children, adults and elderly from 2001 until 2008 in Germany, scaled up by an underreporting rate (cf. Section \ref{complications}) as inferred and  described in more detail in  \cite{wei1}.  Surveillance of rotavirus infection is relevant to public health authorities because it is  the primary cause  for severe gastroenteritis in infants and young children  and  causes significant morbidity, mortality and
financial burdens worldwide \citep{wales}. Individuals can get the disease more than once  but the highest reported incidence is observed in children under the age of 5 and rises again later in life. Parameter estimation from surveillance data is crucial  if one wants to  e.g.\ assess the effect of vaccination campaigns or other  public health control
measures. \\
In \cite{Stocks2017} the disease transmission was modeled as an age-stratified SIRS Markov process with overdispersion in the observation (cf.\ Equation (\ref{nbin})) as well as in the transmission model (cf.\ Equation  (\ref{gammawnoise})). 
As rotavirus in Germany varies seasonally peaking around March, a forcing function (cf.\ Equation (\ref{forcing})) was used. \\
Here, we will present a simplified version of the model which ignores age-structure and perform inference for one realization of this model; the simulated case report data (black line) is shown in Figure \ref{rota}.  The initial values were fixed at the stationary distribution of the system and the waning of immunity rate, $\omega>0$, was also assumed to be fixed and known. With the notation as in Section 
\ref{examplesir}, this translates  mathematically into  
 \begin{align}
\mathbb{P}[\Delta N_{\centerdot S}(t)=1 |\bm{X}(t)]&=  \mu \mathcal{N} \tau +o(\tau)  \nonumber \\
\mathbb{P}[\Delta N_{SI}(t)=1 |\bm{X}(t)]&=  \beta(t) I(t) S(t)\mathcal{N}^{-1}\tau +o(\tau) \quad \nonumber \\
\mathbb{P}[\Delta N_{IR}(t)=1 |\bm{X}(t)]&= \gamma I(t)\tau  +o(\tau) \quad\label{jump_rota}\\
\mathbb{P}[\Delta N_{RS}(t)=1 |\bm{X}(t)]&= \omega R(t)\tau +o(\tau)  \nonumber\\
\mathbb{P}[\Delta N_{A \centerdot}(t)=1 |\bm{X}(t)]&= \mu A(t)\tau  +o(\tau) \quad \text{with $A\in \{S,I, R\}$} \nonumber 
\end{align}
where  $\mu>0$ denotes the birth and death rate which was assumed to be constant and known.  Moreover,  $N_{\centerdot A}(t)$ counts the number of births and $N_{A \centerdot}(t)$  counts the number of deaths with $A \in \mathcal{X}$ in the respective compartment up until time $t$. 
The seasonal-forced transmission rate was defined as 
\begin{equation*}
\beta(t)= \beta \left(1+\rho\cos\left(\frac{2\pi}{w} t +\phi\right)\right) \xi(t), 
\label{forcing}
\end{equation*}
where $\xi(t)$ is defined as in Equation ($\ref{gammawnoise}$).
The transition probabilities from (\ref{jump_rota}) are related to the state variables in the following way:
\begin{align}
\Delta S(t)&= \Delta N_{ \centerdot S}(t)-\Delta N_{ SI}(t) +  \Delta N_{RS}(t)-\Delta N_{S \centerdot}(t) \nonumber\\
\Delta I(t)&= \Delta N_{ SI}(t)-\Delta N_{ IR}(t)-\Delta N_{I \centerdot}(t) \nonumber\\
\Delta R(t)&= \Delta N_{IR}(t)-\Delta N_{RS}(t)-\Delta N_{R \centerdot}(t). \nonumber
\end{align} 
 With this notation 
 the true number of newly infected individuals accumulated  in each observation time period $(t_{n-1},t_{n}]$,   $ n \in \{1,2,\dots,N\}$ is given as 
\begin{equation*}
H(t_n)=N_{SI}(t_{n})-N_{SI}(t_{n-1}).\label{obs_stoc}
\end{equation*} 
\noindent The number of recorded  cases $Y_n$ within this time period was  modeled as
\begin{align*}
Y_n &\sim \text{NegBin}\left( H(t_n), \frac{1}{\psi}\right),
\end{align*}
compare Equation (\ref{nbin}).
In this example we inferred   the susceptibility parameter $\beta$, the seasonal forcing  parameters $\rho$ and $\phi$ and the overdispersion parameters $\sigma$ and $\psi$, hence $\bm{ \theta}=(\beta, \rho,\phi,\sigma, \psi)'$.  A diagnostic plot  similar to the one in Figure \ref{mif} (available online) shows that the iterated filtering algorithm performs well as the  parameter estimates converge towards the parameters used for simulation.   
Figure \ref{rota} shows the model evaluated at the MLE obtained from the iterated filtering. Further implementational details for this example are available at  \cite{theresa_git}. \\
 The aim of the analysis in \cite{Stocks2017} was to compare the impact of different kinds of variabilities in the transmission as well as the observation model. The paper concluded that a model which accounts for overdispersion in both model components (dark and light shading in Figure \ref{rota}) is best suited to explain the analyzed data. 
The strength of using iterated filtering in this context was that the method allowed for efficient maximum likelihood estimation  for a POMP where the underlying  transmission model  was stochastic. Hence, it facilitated model comparison with respect to different choices of variability in both model components.

\section{Concluding comments} 
\label{comments}
\subsection{Pros and cons of iterated filtering methods for outbreak data}
The method we have described has many positive aspects. First of all, as illustrated in this chapter, the method is highly flexible and can be easily adapted to many different situations of interest.
The great strength of the method is that one only needs to simulate from the transmission model if the complete likelihood is not directly tractable. Facilitated by the \texttt{R} package \texttt{pomp}  the method is very straightforward to implement and can be easily adapted for a wide range of settings. The package is  optimized for computational efficiency, in particular the simulation functions (such as \texttt{euler.sim} or \texttt{gillespie.sim})   facilitate implementation considerably and  give rise to   substantial   gains in speed by accessing C code. Secondly, the method facilitates dealing with unobserved variables  and has no problem with the fact that transmission is continuous in time, while observations are made at discrete points in time.  Furthermore, it is easy to combine  transmission noise and observation noise as long as the Markov property is retained.
The method also allows the inclusion of  covariates in mechanistically reasonable ways. 
As mentioned earlier,  to date this is the only simulation-based frequentist approach for POMP models that does not rely on summary statistics  and can perform  efficient  inference for POMP models in situations where available Bayesian methods cannot \citep{Ionides2015}.\\
However, there is one main difficulty with the method. Since it is purely simulation-based it comes at the cost of being computationally very expensive in particular when  the time series becomes long and the state space increases. In Section \ref{tips} we gave some practical hints of how to mitigate the problem. Moreover, the method requires the transmission model to be Markovian which can be problematic because in practice the stages of diseases are rarely well-described by exponential distributions. An easy way to mitigate the problem is by subdividing the respective compartments so that the waiting times individuals stay in each compartment change from being exponentially distributed to gamma distributed, i.e. having a more pronounced mode. 
Sometimes it is also possible to formulate non-Markovian processes as alternate Markov processes with additional state variables (e.g., if $X_n$ depends on both $X_{n-1}$ and $X_{n-2}$, i.e., is not Markov, one could define a new, bivariate Markov process $Z_n = (X_{n},X_{n-1}))$. However, these strategies increase the dimension of the unobserved state space of the transmission model which can result in an increase of Monte Carlo noise in the particle filter. Hence, those strategies are only effective up to the point where it becomes  computationally too expensive to handle this increase of noise.  

\subsection{Further reading}
For the reader interested in more details about the theoretical background of iterated filtering methods we recommend  \cite{ionides06} and  \cite{Ionides2015} for further reading.  
For the reader interested in learning how to work with the \texttt{R} package \texttt{pomp} we recommend the study of  \cite{kingNgu} and to follow some of the very comprehensive online tutorials available through \cite{King2017}. The package  has already been applied to answer many different epidemiological questions for different infectious diseases, e.g.\ for cholera \citep{King2008},  measles \citep{He2010},    malaria \citep{bhadra}, polio \citep{martinezpolio}, ebola \citep{King2015}, and rotavirus \citep{martinez,Stocks2017}.
Finally, iterated filtering has facilitated the development and analysis of a divers array of mechanistic models, a recent and comprehensive review of which can be  found in  \cite{Breto2017}.
\section*{Acknowledgments}
I would like to thank Aaron A. King for making the tutorials and, especially, the codes in \cite{King2017} available. Furthermore, I would like to thank Carles Bret\'{o} and Philip O'Neill for their constructive comments and suggestions. TS was  supported by the Swedish research council, grant number 2015\_05182\_VR.
\bibliographystyle{apalike}
\bibliography{ms}
\end{document}